\documentclass[fleqn,12pt,twoside]{article}
\usepackage{espcrc1}
\usepackage{wrapfig}
\usepackage{graphicx}

\newcommand{\AmS}{{\protect\the\textfont2
  A\kern-.1667em\lower.5ex\hbox{M}\kern-.125emS}}

\title{$J/\Psi \rightarrow ee $ and $J/\Psi \rightarrow \mu\mu$ Measurements 
in $AuAu$ and $pp$ Collisions at $\sqrt{s_{NN}} = 200 \ GeV$}

\author{A. D. Frawley\address[MCSD]{Department of Physics, 
        Florida State University,
        Tallahassee, FL 32306, USA} for the PHENIX Collaboration\footnote{
	for the full PHENIX Collaboration author list and acknowledgements, 
	see Appendix "Collaborations" of this volume.} }
       
\begin{document}

\maketitle

\begin{abstract}

First results are presented from $J/\Psi \rightarrow ee$ and $J/\Psi 
\rightarrow \mu\mu$ measurements by the PHENIX experiment at RHIC. These 
results are from a preliminary analysis of dielectrons from $AuAu$ and $pp$
data and dimuons from $pp$ data measured in the RHIC run 2. A total cross 
section for $J/\Psi$ production in $\sqrt{s} = 
200 \ GeV$ $pp$ collisions of $3.8 \pm 0.6$ (stat) $\pm 1.3$ (sys) $\mu$b 
is found. The $J/\Psi \rightarrow ee$ yield is presented versus number of 
binary collisions.

\end{abstract}

\section{Introduction}

Interesting results obtained at the CERN SPS \cite{Sps1} have 
heightened theoretical interest in $J/\Psi$ production in relativistic heavy
ion collisions. A new generation of models has appeared in which both 
dissociation and 
recombination of $c\bar{c}$ pairs play a significant role. There are now 
model predictions for RHIC central collisions ranging from the almost 
complete disappearance of $J/\Psi$ \cite{Mod1}, to $J/\Psi$ enhancement 
\cite{Mod2}.

In this paper we present the first results from PHENIX on $J/\Psi$ production 
at RHIC.

\section{Measurements and Analysis}

PHENIX \cite{Phnx1} 
detects electrons in the pseudo-rapidity range $\eta = \pm 0.35$, and muons 
in the range $1.2 < \eta < 2.2$. In the central arms, 
electrons are identified by matching 
charged particle tracks to energy deposits in the Electromagnetic Calorimeter 
and rings in the Ring Imaging Cherenkov Detector. 
In the muon arms, muon candidates are found by identifying roads in the 
Muon Identifier, and matching them to tracks found in the Muon Tracker.

During the 2001/2002 RHIC run, PHENIX recorded data from both $AuAu$ 
and $pp$ collisions. Only the south muon arm was operational during
that run.

\subsection{pp}

Data from $pp$ collisions were recorded for an integrated luminosity of 
150 $nb^{-1}$. The minimum bias interaction trigger for the present analysis 
was the Beam-Beam Counter (BBC), with at least one hit per arm. 

Interaction rates for $pp$ varied from about 5 $KHz$ to 30 $KHz$. 
$J/\Psi$ events were selected using level-1 triggers \cite{Lvl1} in coincidence
with the minimum bias interaction trigger. The $J/\Psi 
\rightarrow ee$ trigger used in this analysis 
(about half of our $pp$ data sample)
required that at least 700 MeV be deposited in the Electromagnetic 
Calorimeter. The $J/\Psi \rightarrow \mu\mu$ trigger required at least one 
deep muon and one shallow muon in the Muon Identifier. 
The level-1 triggers allowed all $J/\Psi$ events to be recorded at the 
highest interaction rates. After vertex cuts
($\pm$ 35 cm for $ee$ and $\pm$ 38 cm for $\mu\mu$) and selection of 
good runs, about $1 \times 10^9$ events were sampled for $ee$ and $1.7 
\times 10^9$ events were sampled for $\mu\mu$. The muon arm data were 
subdivided into two rapidity bins.

Detector acceptance, level-1 trigger efficiency, and detector efficiency 
were estimated for a typical run
using simulations of single $J/\Psi$ decays with a realistic detector response.
An average correction for run-to-run variations in detector 
active area was then determined from the azimuthal distribution of 
electrons/event for all runs used in the analysis.

For the $ee$ case, two models of the transverse momentum distribution and a 
range of values for the mean p$_T$ 
were used to establish the systematic uncertainty in the acceptance.
This was necessary because the p$_T$ distribution is not well defined by the 
data due to low statistics. This systematic uncertainty was estimated to be
10\%.

For both the $ee$ and $\mu\mu$ cases,  
PYTHIA simulations were used to estimate the BBC trigger efficiency 
for inelastic $pp$ events (needed to calculate the total number of events),
and to estimate the fraction of events containing a $J/\Psi$ that fired the 
BBC trigger (ie. the BBC trigger efficiency for $J/\Psi$ detection).

For cross section calculations, the RHIC luminosity was assumed to 
have a systematic uncertainty of 20\% in this preliminary analysis. This 
uncertainty is expected to be reduced considerably with more study.

Systematic errors were added in quadrature to produce an
overall systematic uncertainty of 35\% for $pp \rightarrow J/\Psi 
\rightarrow ee$ and 29\% for $pp \rightarrow J/\Psi \rightarrow 
\mu\mu$ cross section data.

\subsection{AuAu}

Data from $AuAu$ collisions were recorded 
for an integrated luminosity of 24 ${\mu}b^{-1}$. 
The minimum bias trigger used for $AuAu$ collisions was a coincidence
between the BBC (2 hits per arm) and the Zero Degree Calorimeter (ZDC). 
At the highest 
luminosities, where only a fraction of the interactions could be recorded, 
level-2 triggers \cite{Lvl1} were used to select $J/\Psi$ events and minimum 
bias triggers were allocated only $\sim$ 20\% of the data bandwidth.
Interaction rates varied from about 100 $Hz$ to 1600 $Hz$. At the end 
of the $AuAu$ run, 
the data samples from minimum bias triggers and level-2 triggers were
approximately equal.

In the present preliminary analysis of the $AuAu$ data, only 
the $J/\Psi \rightarrow ee$ data have been analyzed, and only minimum bias 
triggers were used. After applying a vertex cut of $\pm 30$ cm and 
selecting good runs, $\sim$ 26 million events were analyzed.

Yields were extracted using 7 different procedures for yield and background 
estimation, to establish the systematic errors. The acceptance and 
detector efficiency were estimated using simulations of
single $J/\Psi$ decays with a realistic detector response. The minimum bias
trigger is estimated \cite{Mb1} to have unit efficiency for the most central 
90\% of interactions used in this analysis. 

The event centrality was estimated \cite{Mb1} using the combined data from 
the BBC and ZDC detectors. The data were separated into three centrality 
bins. The centrality dependence of the efficiency for offline $J/\Psi$
reconstruction was estimated by embedding simulated single electrons
into measured events and then taking the square of the single electron 
reconstruction efficiency as the $J/\Psi$ reconstruction efficiency due to
occupancy.

Combining systematic errors quadratically produced
an overall systematic uncertainty per centrality bin of 26\% (0-20\%) 
27\%(20-40\%) and 26\%(40-90\%).

\section{Results}
The dielectron invariant mass spectrum from $AuAu$ (minimum bias) is shown 
in Fig. 1, and that from $pp$ is shown in Fig. 2. Fig. 3 shows the 
$pp$ dimuon invariant mass spectrum. 

\begin{figure}
\begin{minipage}[b]{0.33\linewidth}
\centering
\includegraphics[width=1.0\textwidth]{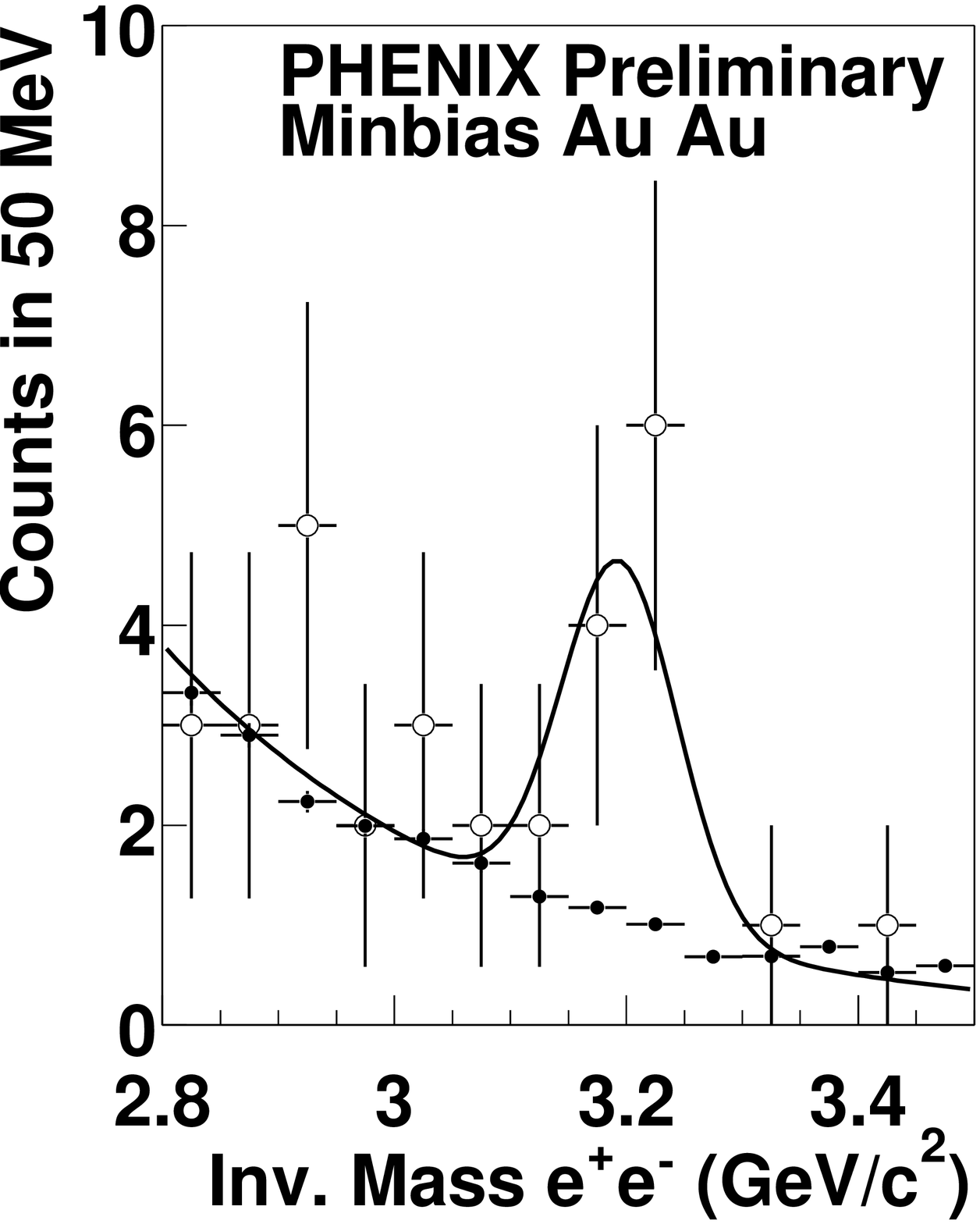}
\end{minipage}%
\begin{minipage}[b]{0.33\linewidth}
\centering
\includegraphics[width=1.0\textwidth]{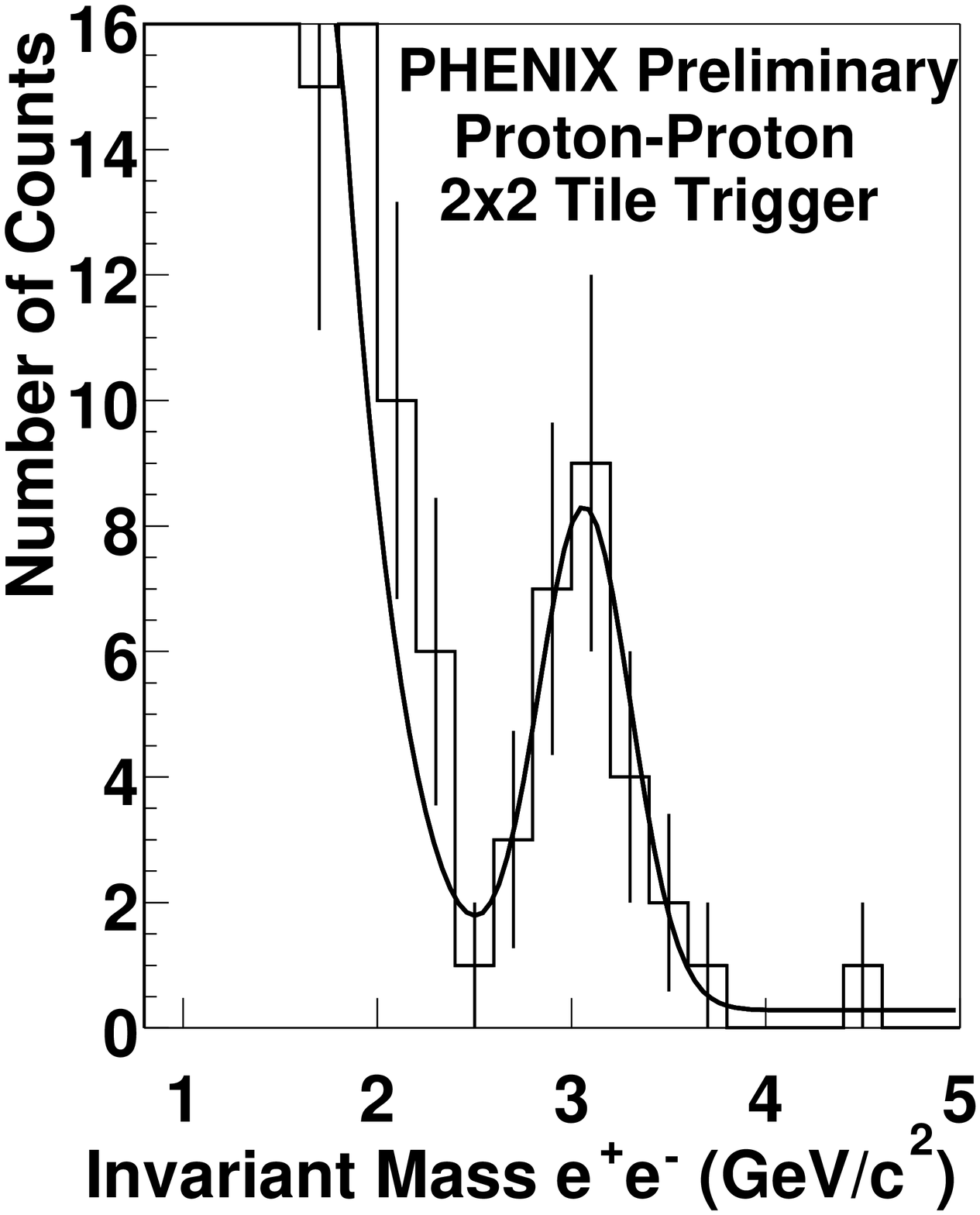}
\end{minipage}%
\begin{minipage}[b]{0.33\linewidth}
\centering
\includegraphics[width=1.0\textwidth]{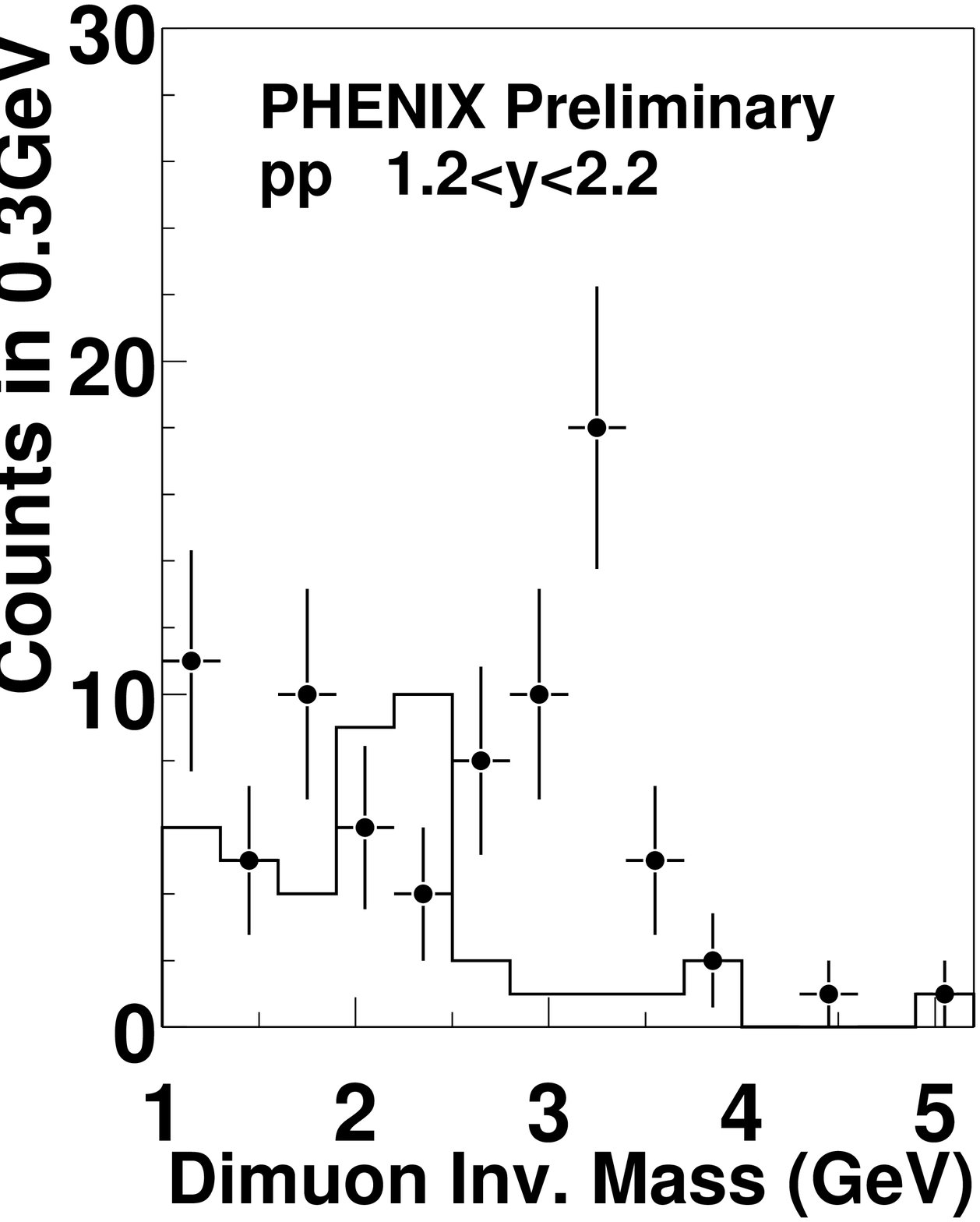}
\end{minipage}\\
\begin{minipage}[t]{0.3\linewidth}
\centering
\caption{Minbias $AuAu \rightarrow ee$ invariant mass spectrum, 
with mixed background and gaussian plus exponential fit to the $J/\Psi$ peak.}
\end{minipage}%
\hspace{0.2in}
\begin{minipage}[t]{0.3\linewidth}
\centering
\caption{Minbias $pp \rightarrow ee$ invariant mass spectrum, 
with gaussian fit to the $J/\Psi$ peak.}
\end{minipage}%
\hspace{0.2in}
\begin{minipage}[t]{0.3\linewidth}
\centering
\caption{Minbias $pp \rightarrow \mu\mu$ invariant mass spectrum (points), 
with unlike-sign pair background (histogram).}
\end{minipage}
\end{figure}

The $J/\Psi$ rapidity distribution was obtained by combining the data from the 
$ee$ and $\mu\mu$ channels, and is shown in Fig. 4. The total cross section
was obtained by fitting both a Gaussian and the PYTHIA prediction for 
the shape of the rapidity distribution, using the parton distribution function
GRV94LO. Both gave essentially the same 
integral, and the average of the two results for the total cross section is
3.8 $\pm$ 0.6 (stat) $\pm$ 1.3 (sys) $\mu$b. This value agrees well with the 
prediction of the Color Evaporation Model \cite{Ce1}.

\begin{figure}
\begin{minipage}[b]{0.5\linewidth}
\centering
\includegraphics[width=1.0\textwidth]{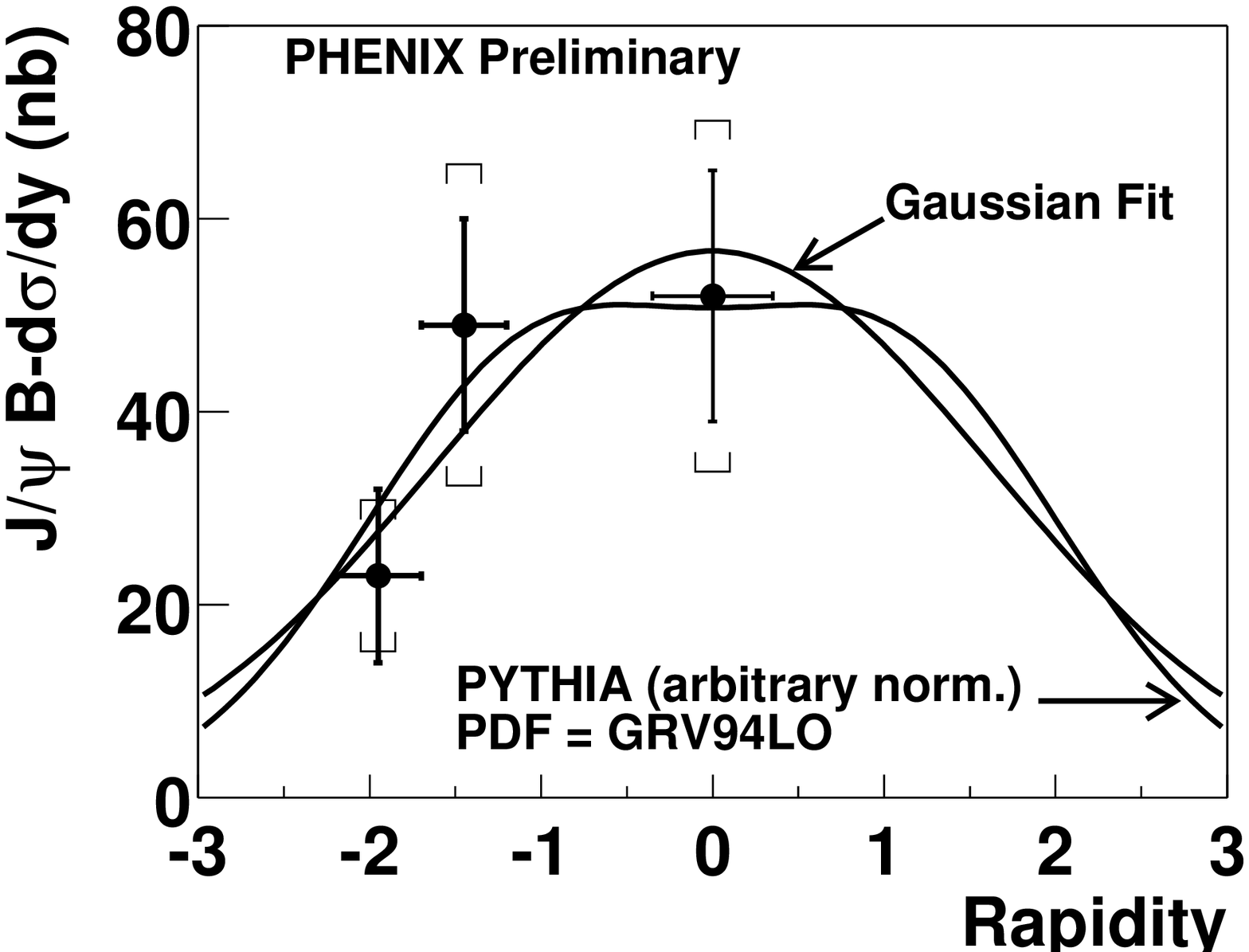}
\end{minipage}
\begin{minipage}[b]{0.5\linewidth}
\centering
\includegraphics[width=1.0\textwidth]{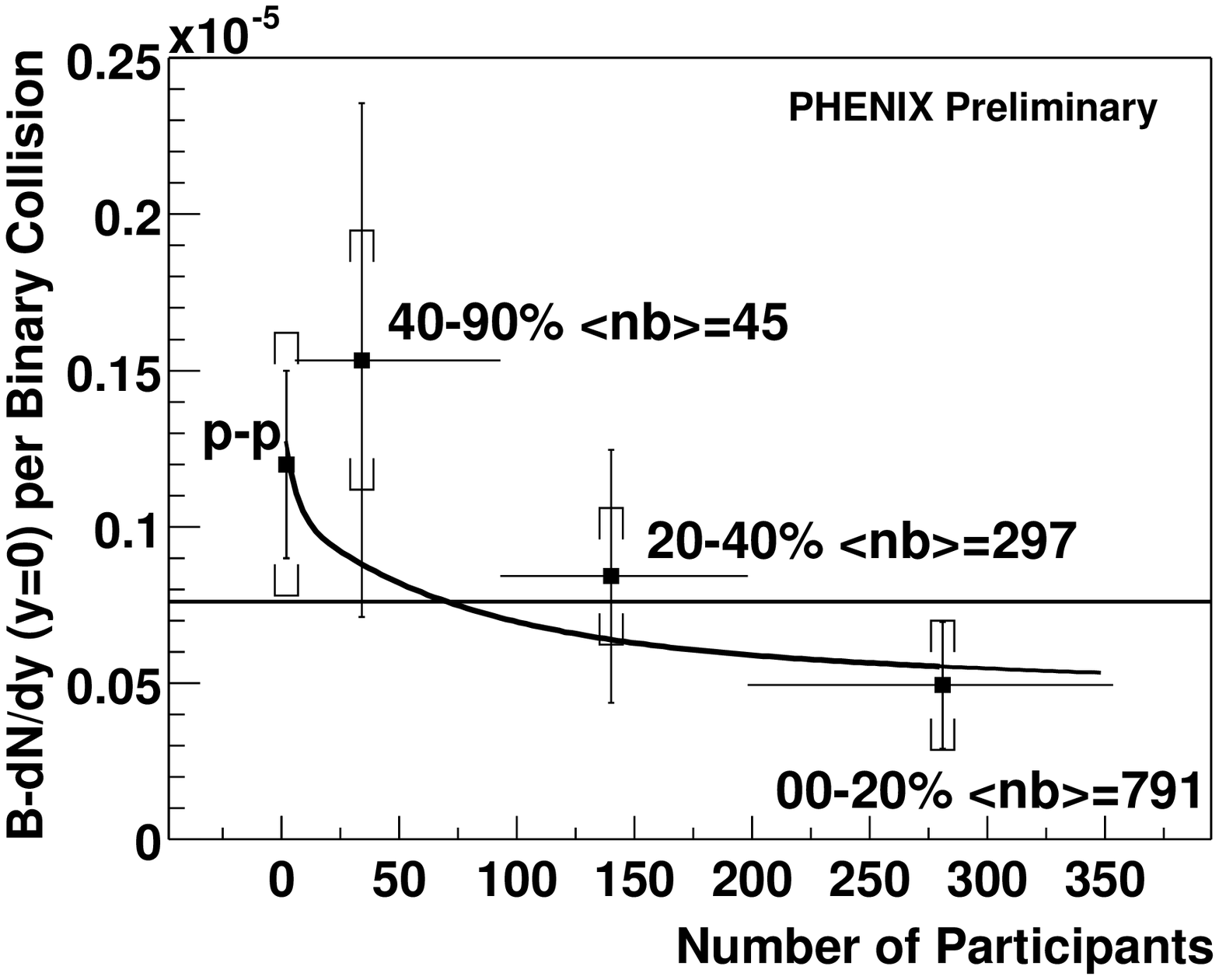}
\end{minipage}
\begin{minipage}[t]{0.45\linewidth}
\caption{$J/\Psi$ branching ratio times dN/dy vs rapidity for $pp$ 
Brackets are maximum plausible systematic spreads.} 
\end{minipage}
\hspace{0.5in}
\begin{minipage}[t]{0.45\linewidth}
\caption{$J/\Psi \rightarrow ee$ branching ratio times dN/dy scaled by 
N$_{binary}$. The 
flat line is the best fit binary scaling value. The curve is 
a normal nuclear absorption model calculation \cite{Abs1}.}
\end{minipage}
\end{figure}

The $J/\Psi \rightarrow ee$ yield versus the number of participants 
(N$_{part}$) is shown 
in Fig. 5 for the $pp$ data and for the $AuAu$ data divided 
into three centrality bins. 

\section{Discussion and conclusions}

Rather limited conclusions can be drawn from the data in Fig. 5 
because of the large statistical and systematic uncertainties associated 
with this preliminary analysis. We have fitted 
a flat line to the data to see if the data are consistent with binary scaling.
The best fit, shown in Fig. 5, has a confidence level of 16\% based on 
the statistical uncertainties alone. Thus the binary scaled $J/\Psi$ yields 
probably trend down with increasing N$_{part}$, but that is not a 
strong statement. 

We also show in Fig. 5 a comparison of the N$_{part}$ dependence with a 
calculation from a simple normal nuclear
absorption model assuming a 7.1 mb absorption cross section \cite{Abs1}. 
Here the confidence level is found to be 80\%. However we wish to stress 
that the uncertainties in the data are too large to allow discrimination 
between any but the most extreme scenarios. 

We expect to approximately double the number of measured $J/\Psi \rightarrow 
ee$ when we add level-2 triggered events to this analysis. We also hope to
extract a comparable sample of $J/\Psi \rightarrow \mu\mu$ data at forward 
rapidity from the existing data set.

\end{document}